\documentclass[conference]{IEEEtran}
\IEEEoverridecommandlockouts

\usepackage{cite}
\usepackage{amsmath,amssymb,amsfonts}
\usepackage[mathscr]{euscript}
\usepackage{graphicx}
\usepackage{textcomp}
\usepackage{xcolor}
\usepackage{amsthm}
\usepackage{cancel}
\usepackage{comment}
\usepackage{lipsum}
\usepackage{hyperref}
\usepackage{subfig}
\usepackage{placeins}
\usepackage{listings}
\usepackage[capitalize]{cleveref}
\usepackage[table]{xcolor}
\usepackage{colortbl}

\definecolor{codegray}{rgb}{0.5,0.5,0.5}
\definecolor{codepurple}{rgb}{0.58,0,0.82}
\definecolor{backcolour}{rgb}{0.95,0.95,0.92}

\lstdefinestyle{pythonstyle}{
    backgroundcolor=\color{backcolour},
    commentstyle=\color{codegray}\ttfamily,
    keywordstyle=\color{blue},
    numberstyle=\tiny\color{codegray},
    stringstyle=\color{codepurple},
    basicstyle=\ttfamily\footnotesize,
    breakatwhitespace=false,
    breaklines=true,
    captionpos=b,
    keepspaces=true,
    numbers=left,
    numbersep=5pt,
    showspaces=false,
    showstringspaces=false,
    showtabs=false,
    tabsize=2,
    language=Python
}

\newcommand{\dram}[0]{\mathscr{S}_{3}}
\newcommand{\sram}[1]{\mathscr{S}_{#1}}
\newcommand{\size}[1]{sz_{#1}}

\newcommand{\CORE}[1]{c_{#1}}
\newcommand{\BANK}[1]{b_{#1}}

\definecolor{headergray}{RGB}{230,230,230}
\definecolor{footergray}{RGB}{210,210,210}

\def\BibTeX{{\rm B\kern-.05em{\sc i\kern-.025em b}\kern-.08em
    T\kern-.1667em\lower.7ex\hbox{E}\kern-.125emX}}
\begin{document}

\title{ALADIN: \underline{A}ccuracy–\underline{L}atency–\underline{A}ware \underline{D}esign-Space \underline{I}nfere\underline{N}ce Analysis for Real-Time \\Embedded AI Accelerators}

\author{
\IEEEauthorblockN{Tommaso Baldi\IEEEauthorrefmark{1},
Daniel Casini\IEEEauthorrefmark{1},
Alessandro Biondi\IEEEauthorrefmark{1}}
\IEEEauthorblockA{\IEEEauthorrefmark{1}
Department of Excellence in Robotics and AI, TeCIP\\
Scuola Superiore Sant'Anna\\
Pisa, Italy\\
\{tommaso.baldi, daniel.casini, alessandro.biondi\}@santannapisa.it}
}

\newcommand\remove[1]{{\color{red} #1}}
\newcommand\new[1]{{\color{blue}#1}}

\newcommand\TODO[1]{{\color{red} TODO: #1}}

\maketitle

\begin{abstract}
The inference of deep neural networks (DNNs) on resource-constrained embedded systems introduces non-trivial trade-offs among model accuracy, computational latency, and hardware limitations, particularly when real-time constraints must be satisfied. This paper presents ALADIN, an accuracy–latency–aware design-space inference analysis framework for mixed-precision quantized neural networks (QNNs) targeting scratchpad-based AI accelerators. ALADIN enables the evaluation and analysis of inference bottlenecks and design trade-offs across accuracy, latency, and resource consumption without requiring deployment on the target platform, thereby significantly reducing development time and cost.

The framework introduces a progressive refinement process that transforms a canonical QONNX model into platform-aware representations by integrating both platform-independent implementation details and hardware-specific characteristics. ALADIN is validated using a cycle-accurate simulator of a RISC-V–based platform specialized for AI workloads, demonstrating its effectiveness as a tool for quantitative inference analysis and hardware–software co-design. Experimental results highlight how architectural decisions and mixed-precision quantization strategies impact accuracy, latency, and resource usage, and show that these effects can be precisely evaluated and compared using ALADIN, while also revealing subtle optimization tensions.
\end{abstract}


\section{Introduction}
\label{sec:intro} 

The convergence of Artificial Intelligence (AI) and the Internet of Things (IoT) is reshaping how we interact with and interpret the world around us. IoT devices are traditionally embedded systems designed to be lightweight, cost-effective, and energy-efficient. However, the rising complexity of AI models is pushing the boundaries of what these resource-constrained devices can handle~\cite{IIoT}. 
As a result, there is an increasing need to rethink the hardware and software architecture of resource-constrained systems to support more demanding AI workloads~\cite{silvano2023survey}. This growing resource demand is driving innovation in edge computing, model optimization, and specialized hardware accelerators. Techniques like pruning~\cite{cheng2024survey}, quantization~\cite{rokh2023comprehensive}, and knowledge distillation~\cite{gou2021knowledge} are being employed to compress AI models, making them more suitable for deployment on constrained devices.

The tradeoff between Deep Learning (DL) algorithm accuracy and the resources available on the device becomes even more complex when we operate under strict real-time constraints, where low-latency and predictable responses are critical~\cite{kang2019deeprt}. Examples of such time-critical systems are autonomous vehicles, industrial control systems, medical monitoring devices, and robotics.

Ensuring a timing-predictable behavior while executing non-trivial AI workloads is particularly challenging on heterogeneous, resource-constrained platforms where processing, memory, and energy resources are limited~\cite{singh2023edge}. 
Handling this tradeoff is not trivial and may lead to many complex and time-consuming design-exploration iterations. 

However, exploring these accuracy–latency–resource trade-offs directly on real platforms typically requires implementing and porting each candidate model configuration to the target hardware, followed by time-consuming deployment and evaluation cycles. Furthermore, this results in a trial-and-error approach that severely limits the portion of the design space that can be evaluated in practice, as only a small number of configurations can be evaluated within realistic development budgets. As a consequence, hardware–software (HW-SW) co-design decisions, such as the sizing of memories and computing units are often taken late in the design flow and with limited quantitative insight, hindering the ability to systematically optimize AI workloads under real-time constraints.


\smallskip
\noindent \textbf{This paper.} In this paper, we propose a workflow to evaluate and explain the samples coming from candidate mixed-precision and implementation configurations, possibly generated by external automated design-space exploration methods~\cite{automl, ANNA, Quantune, QuantuneV2}, allowing users to uncover possible bottlenecks during inference, and, when possible, also guide the design of the hardware (HW) in scratchpad-based heterogeneous AI accelerators. To this end, we employ model refinements and scheduling strategies that explicitly consider practical implementations of DNN operations and the underlying hardware platform, particularly focusing on available on-chip SRAM and dedicated accelerators. 
By starting from a canonical QONNX model of quantized DNNs, we decorate the model to incorporate information related to the implementation of the layer operations, such as the amount of required compute operations and data in memory.
A further refinement is eventually applied to split operations into sub-operations that can individually be scheduled on the target hardware platform. To do so, we consider a general, yet abstract model of a scratchpad-based AI accelerator.

The resulting design evaluation framework allows exploring trade-offs between latency, accuracy (which depends on the selected quantization scheme), and required hardware resources by working on models only, i.e., reducing the number of trial-and-error compilation and execution of QNNs on the target hardware. 
ALADIN outputs the inference latency experienced by a model inference instance, which can be compared with its deadline to assess the satisfaction of real-time constraints. In this way, ALADIN enables the screening of candidate quantization and implementation configuration based on deadline feasibility.
Finally, we evaluate the workflow on GVSoC~\cite{gvsoc}, a cycle-accurate RISC-V-based AI platform, showing how our approach enables HW–SW co-design and highlights the impact of different design choices.

\section{Background}
This section reviews the essential background on quantized neural networks and look-up table implementations.

\subsection{Quantized Neural Networks}
\label{subsec:backgroundquantized}

Quantization~\cite{survey_quant, survey_quant2}, is one of the most widely used techniques adopted for porting Neural Networks (NNs) models on highly resource-constrained devices. Quantization can be seen as the mapping from input values in a large (often continuous) set to output values in a small (and finite) set. Efficient representation of numerical values is important in NNs, which are heavily over-parameterized \cite{int_vs_fp}, making it possible to significantly reduce bit precision while preserving accuracy. Quantization can be divided into two main categories: uniform and non-uniform quantization.

\smallskip
\noindent
\textbf{Uniform quantization.} It consists of an affine transformation where each value is mapped to an integer value and all bins\footnote{In the quantization jargon, a bin refers to the interval of continuous input values that are all mapped to the same quantized (integer) level.} have the same width. The transformation can be defined as:
    \begin{equation}\label{eq:quant}
        Q(r) = Int(r/S) - Z,
    \end{equation}
where $Int()$ denotes rounding clipping (details will follow), $r$ is the full precision representation of the value to be quantized, and $Z \in \mathbb{Z}$ is the shift component that is used to handle asymmetric distributions (i.e., not centered around zero). $S \in \mathbb{R}$ is the scale factor, and it is computed as $S = (\beta - \alpha) / (2^B - 1)$,
where $B$ is the target bit-width, while both $\alpha$ and $\beta$ are the boundaries of the representations (i.e., the minimum and maximum values that we expect to represent
(different methods for computing them are available~\cite{quant_white}). For convolutional layers, it is also possible to reduce the performance degradation by adopting channel-wise quantization-each out channel of the convolution has its own quantization configuration ($S$ and $Z$), at the cost of a higher memory footprint since quantization parameters are not anymore scalars~\cite{quant_white}.

The $Int()$ operation in Equation \eqref{eq:quant} denotes the concatenation of rounding and clipping (which ensures that the value lies within the specified range) between the representation boundaries, which depend on the bit-width and representation scheme (signed or unsigned). Based on the implementation, the rounding can be performed using different implementations (e.g.,  \texttt{round}, \texttt{floor}, \texttt{ceil}). Typically, uniform quantization is widely adopted due to its mapping characteristics, which makes it less expensive to be implemented in hardware: i.e., quantization can be implemented in full integer logic by applying dyalic scaling~\cite{jacob2018quantization}.

\smallskip
\noindent
\textbf{Non-uniform quantization.} This quantization strategy is more flexible with respect to uniform quantization, as it allows variable bin widths that can be tailored to the actual data distribution. Indeed, it is characterized by a transformation where the mapping bins have an arbitrary width, which can be tuned, for instance, to give more precision to values closer to zero~\cite{apot}, or when a specific layer is more sensitive then other~\cite{SEMQ}. The transformation can be defined as
$        Q(r) = x_i \text{, if } r\in [\Delta_i, \Delta_{i+1}) $
where $r$ is the full precision representation of the value to be quantized, $x_i \in \mathbb{Z}$ is the corresponding integer representation, and $\Delta_i \in \mathbb{R}$ for $i = 1, \dots, B$ are the boundaries between bins.

\subsection{Look-up tables} \label{sec:lut}
It has been shown that the performance of typical QNN operations, such as matrix multiplications between matrices composed of integer values, can be improved by pre-computing all possible partial products in a look-up table (LUT) \cite{lut, lut2}. The LUT stores all the pre-computed products between all possible combinations of values that the inputs may assume. Then, instead of applying the Multiplication-Accumulation (MAC) operation, we just need to access the LUT using the weights and the activations as indices of the table and accumulate the partial product in a dedicated register. They hence allow \emph{trading computing demand by memory space requirements}.

This approach can also leverage the parallelism of the architecture since more cores in parallel can read the entries of the LUT, boosting the computation time. In terms of memory, the amount of required parameters increases, and the dimension of the LUT can be computed as
 $2^{L_w + L_a} \cdot L_{acc}$,
where $L_w$ and $L_a$ are respectively the fixed bit-width of the quantized weight and input, while $L_{acc}$ is the bit-width of the accumulator. The computation speed-up can be bounded by the memory capacity of the platform, since based on the specific quantization configurations, the dimension of the table may not fit anymore in a local L1 scratchpad (a similar configuration is envisioned in several works, e.g.,~\cite{XpulpNN}, requiring expensive DMA requests to swap data). Our framework allows understanding which are the most convenient implementation options for a mixed-precision quantized network.

\section{Related Work}
The literature on design optimization of accelerated neural network is very vast.
We hence believe appropriate focusing on the works that are close in scope to this paper, which mainly fall into two categories: real-time inference of neural networks and mixed-precision quantization.

One of the first works on the real-time inference of neural networks is DeepRT~\cite{kang2019deeprt}, a framework supporting a more predictable inference on GPU by dynamically minimizing the tardiness with device-specific tuning knobs. While the framework leverages model compression~\cite{han2015deep}, it does not support design evaluation of different configurations with mixed-precision quantization.

Most of the works targeted real-time inference of deep networks on GPU, FPGA, or CPU platforms~\cite{bateni2018apnet, bateni2020co, zhou2018s, casini2020timing, bateni2018predjoule, ji2022demand, FRED}. 
For example, Shirvani~\cite{shirvani2024duojoule} recently introduced DuoJoule, a framework to address the challenge of meeting latency deadlines and energy constraints in GPU-based frameworks. Other works targeted the problem of real-time image offloading to servers~\cite{wang2023progressive}. 

Overall, from a real-time scheduling perspective, most of the attention has been attributed to either GPU or CPU platforms.

Quantization is intensively adopted in specialized accelerators (see~\cite{silvano2023survey} for a comprehensive survey), but, to the best of our knowledge, no previous work considered the design-time inference evaluation of implementation strategies under mixed-precision quantization with focus on real-time timing constraints, since most of the works consider general performance constraints such as throughput. Concerning more specialized accelerators and mixed-precision quantization, several relevant works in the literature~\cite{automl, hawq, paretoQ} have demonstrated the potential of mixed-precision quantization, particularly for deep neural networks (DNNs), which are often too large and computationally intensive to be deployed on embedded systems in their full-precision form after training.

The key idea behind mixed-precision quantization is to adopt different quantization strategies on a block- or layer-wise basis. State-of-the-art (SoTA) research has shown that certain layers are more sensitive to quantization than others, motivating the development of adaptive, heterogeneous precision strategies across the network. However, in layer-wise mixed-precision quantization, the design space grows exponentially with the number of layers. Specifically, if $L$ is the number of layers and $B$ is the number of available bit-widths, the total number of possible quantization configurations is $B^L$. This exponential growth necessitates the use of efficient techniques to enable the evaluation of the design-space.


For example, the work in~\cite{automl} frames the search problem  as an RL problem, but its exploration becomes very time-consuming on deep networks. In contrast,~\cite{hawq} uses the Hessian trace to estimate each layer’s sensitivity and allocate higher bit-widths accordingly, leading to faster exploration with comparable performance to~\cite{automl}, despite the non-trivial cost of computing the Hessian trace.

In this work, we focus on QNNs with mixed-precision integer quantization and evaluate the impact of different quantization schemes not only on classification performance but also on inference latency, which was not explored by the aforementioned works. This is particularly relevant because the precision of each layer significantly affects the volume of data that must be transferred to memory, and consequently, the overall computational cost.




\section{System Model}

\begin{figure}[]
    \centering
    \includegraphics[width=0.9\linewidth]{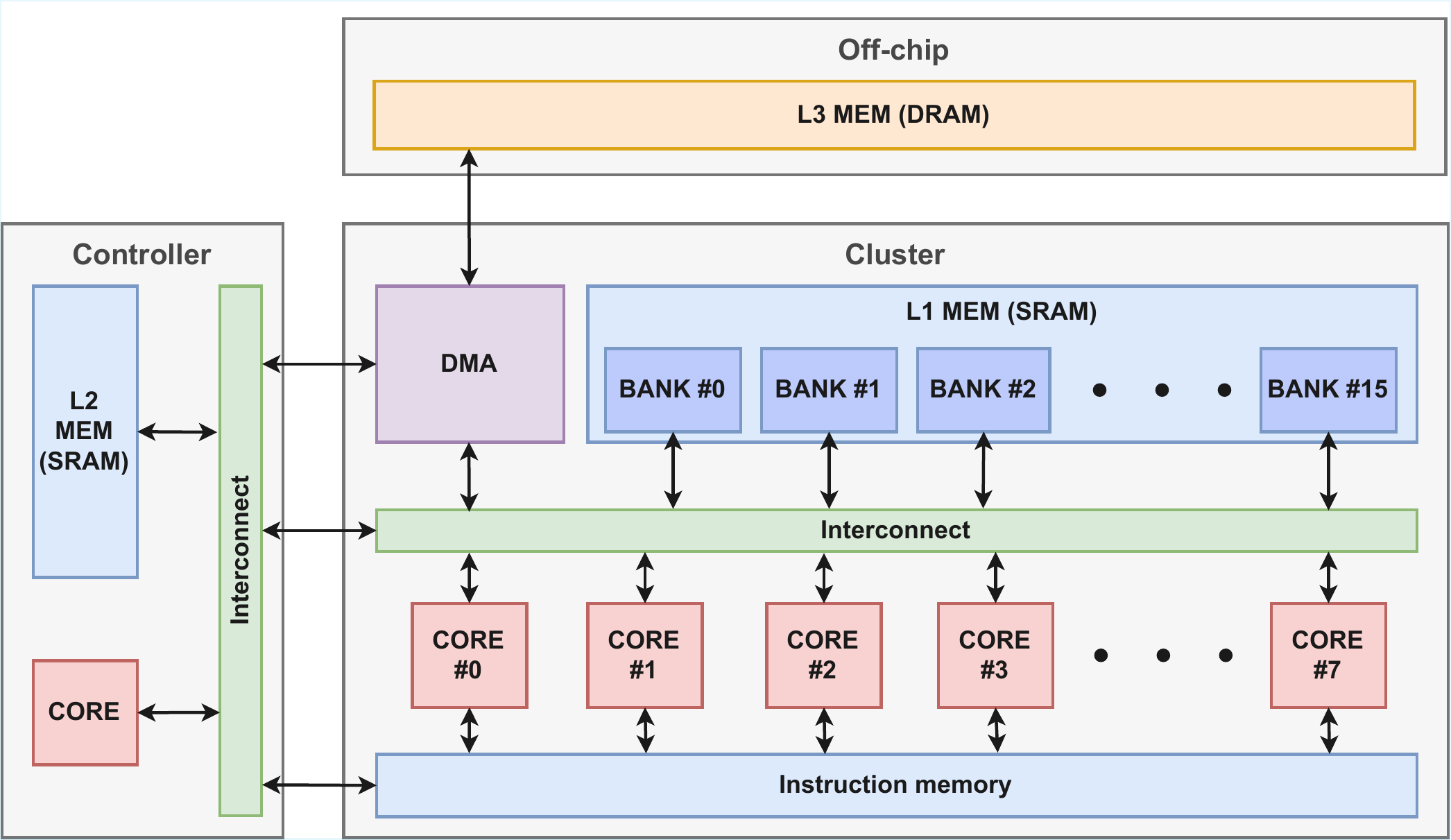}
    \caption{Simplified diagram of a platform with a controller and a parallel cluster with 8 cores and 16 memory banks.}
    \label{fig:platform}
\end{figure}

\begin{figure}[]
    \centering
    \includegraphics[width=1\linewidth]{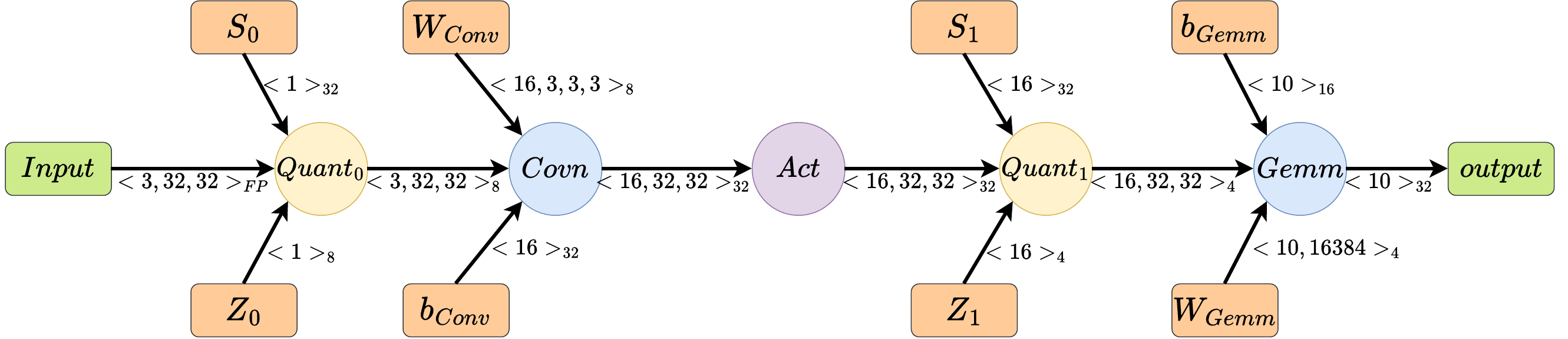}
    \caption{Example of a DAG representation of a simple CNN consisting of a 2D convolutional layer and a fully connected layer, with a mixed-precision quantization configuration. Circles denote the operation \textit{nodes} in the DAG, while squares label the \textit{edges}, representing data dependencies.}
    \label{fig:DAG}
\end{figure}

The platform model considered in this paper is inspired by the main characteristics of modern AI accelerators. Instead of focusing on highly custom accelerators designed for specific use case, we preferred to focus on more general-purpose AI oriented chips, namely the STM32N6 series, equipped with the ARM Cortex M-55 core \cite{arm-cortex}, and GAP8~\cite{GAP8} equipped with RISC-V cores with an extended ISA designed to boost AI operations \cite{XpulpNN}, to foster the generality and applicability of our research. Both chips are characterized by at least two levels of memory hierarchy, composed of a small L1 memory and a bigger-but-slower on-chip L2 memory (both implemented as a scratchpad  memory). 
Such architecture is particularly suitable for time-predictable inference, especially when compared with cache-based architectures with hard-to-predict cache eviction and management policies~\cite{cache_survey, cache_survey_2}.
An off-chip L3 DRAM that provides high capacity, but at a slower speed. These composite memory hierarchies are required due to the increasing memory footprint of modern DNN inference tasks. The data movement among these memories requires explicit DMA transfers and cache locking to avoid cache evictions and allow the memory to behave like a scratchpad, characterized by fixed content that can be controlled programmatically. 

\subsection{Platform model}
\label{sec:platformmodel}

The considered computing platform is composed of a controller part and a core cluster part, as shown in Fig.~\ref{fig:platform}. The cluster is composed of $M$ identical cores $\CORE{1}, \dots,\CORE{M}$. The cores of the cluster share a L1 memory $\sram{1}$ divided in $N$ equal sized banks $\BANK{1}, \dots,\BANK{N}$. Each bank can be accessed in a contention-free manner by at most one device (core or DMA) at a time. The sizes of $\sram{1}$ and $\sram{2}$ are respectively $\size{1}$ and $\size{2}$.
The controller part comprises one core and a local scratchpad memory $\sram{2}$. The core of the controller is responsible for the orchestration of the platform, specifically it offloads the intensive computation of AI-oriented operations to the cluster and it coordinates the data movements between L3-L2 memories and between L2-L1 memories.

A global L3 memory $\dram$ is also present and it is accessible only from the core of the controller, which we assume to be always large enough to store the required data. Indeed, usually its size is way larger with respect to the on-chip memories, i.e., $\size{\dram} >> \size{2}$. Memory sizes are expressed in \textit{chunks}, where each chunk is composed of a given number of bytes. 

\subsection{Application model}
\label{sec:applicationmodel}

The inference of a QNN is based on the neural network architecture and its corresponding quantization configuration. We assume the QNN has been finetuned with one of the available frameworks, such as Brevitas~\cite{brevitas} or QKeras~\cite{qkeras}, using any mixed-precision scheme, and exported to QONNX~\cite{qonnx}.
QONNX
~\cite{qonnx} is an open-source extension of the widely adopted ONNX~\cite{onnx} format. ONNX provides an extensible computation graph model, including standard operator definitions and data types, and it has broad support across machine learning frameworks. QONNX enhances ONNX by introducing support for arbitrary-precision uniform quantization. This added flexibility allows to efficiently handle low-precision quantization setups and extract detailed insights regarding both the input/output data requirements and the complexity of each operation in the graph. Additionally, QONNX is easily extensible in incorporating custom operations and implementations.

A QNN in QONNX is represented as a directed acyclic graph (DAG), as illustrated in Figure~\ref{fig:DAG}. In this representation, circles denote operation nodes, each corresponding to a fundamental QNN operation. Common node types include:

\begin{itemize}
    \item \textit{Quantization (\texttt{Quant})}: These nodes perform re-quantization of inputs or intermediate feature maps. After arithmetic operations, feature maps typically increase in precision (e.g., 16- or 32-bit to prevent overflow), necessitating conversion back to the target precision. Quantization nodes require the associated parameters, i.e., uniform scales $S$ and zero-points $Z$, or threshold sets for non-uniform schemes, to execute this transformation.
    
    \item \textit{2D Convolution (\texttt{Conv})}: The core operation in convolutional neural networks (CNNs) that extracts spatial features from input tensors. Convolution nodes accept quantized weights $W$ and biases $b$ as parameters.
    
    \item \textit{General Matrix Multiplication (\texttt{Gemm})}: A fundamental operation in fully connected layers, especially within a CNN’s classification head. Gemm nodes likewise consume quantized weight $W$ and bias $b$ tensors.
    
    \item \textit{Activation (\texttt{Act})}: Nonlinear functions applied to feature maps following linear operations. Activation nodes do not require learnable parameters.
\end{itemize}

Formally, the DAG is defined as $\mathcal{G} = (\mathcal{V}, \mathcal{E})$, where $\mathcal{V} = \{v_i \mid i = 1, ..., k\}$ is the set of $k$ nodes, each representing an operation performed on the input data to generate the output, and $\mathcal{E} = \{e_{ij} \mid v_i, v_j \in \mathcal{V}\}$ is the set of edges denoting data dependencies between operations. Data is represented as a tensor $<x_1, \ldots, x_n>_b$, where $x_1, \ldots, x_n$ are the tensor dimensions, and $b$ denotes the bit-width of each tensor element. The QONNX format is a fundamental building block of our framework, which we use to extend the QNN representation with implementation-related and platform-related information.



\section{Methodology}
\label{sec:method}

\begin{figure}[]
    \centering
    \includegraphics[width=1\linewidth]{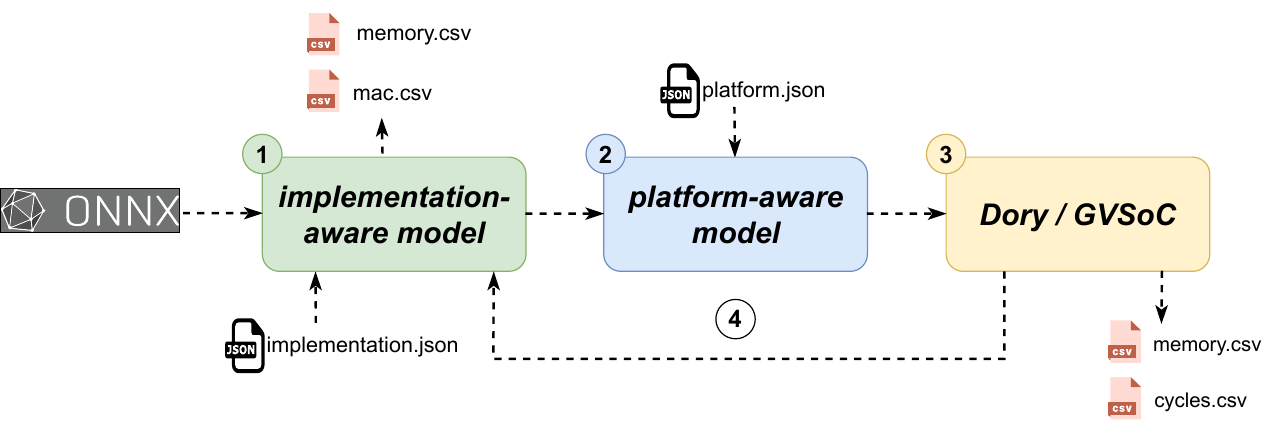}
    \caption{Design inference analysis workflow.}
    \label{fig:design}
\end{figure}

This section presents the proposed workflow, which uses model refinements to eventually produce a detailed model that can enable scheduling while also carrying both implementation- and platform-dependent information.
Model refinements are introduced next and later defined in detail in the following subsections.
Figure~\ref{fig:design} highlights the workflow, which is organized into three main steps:
\begin{enumerate}
    \item \textbf{Generating the \textit{implementation-aware model}}. The workflow starts with a QNN represented in QONNX format.
    Aiming at understanding the tradeoff between model accuracy, latency, and required hardware resources, we enrich and specialize the QONNX model by adding implementation-related details, such as how activation functions or operations are implemented. For example, storing results in a look-up table vs. performing the actual calculation requires a different amount of resources and will likely imply different latencies. Furthermore, mixed-quantization strategies have an impact on both accuracy, hardware resources, and latency. The resulting decorated model, which we call \textit{implementation-aware model}, still abstracts from the actual platform used for deployment, fostering the generality of the procedure.
    \item \textbf{From \textit{implementation\-aware model} to \textit{platform-aware model}}. To actually explore the effects on latency and the feasibility of the deployment on a target hardware platform, a further specialization is required. At this stage, the platform specification (in the form of a hardware model introduced above), is used to further refine 
    the implementation-aware model from the previous step into a \textit{platform-aware model} of the QNN. The latter 
    contains nodes representing operations with a granularity level that is sufficient to generate a schedule on the target hardware while guaranteeing memory-space feasibility to access the L1 scratchpad.
    \item \textbf{Dory and GVSoC}. Here, we use the platform-aware QNN model as input to Dory~\cite{dory}, a C-code generator for MCU platforms. We extended Dory to process our DAG and generate optimized C code for the target platform. The generated code is then compiled and executed on GVSoC~\cite{gvsoc}, a cycle-accurate simulator of RISC-V–based SoCs. From execution traces, we extract detailed information on memory utilization (L1 and L2) and layer-wise clock-cycle requirements for all operations.
    \item \textbf{Closing the loop}. The preceding three steps allow bounding inference latency while retaining information about the adopted quantization scheme (from the input QONNX model) and the corresponding implementation choices and hardware resources on which the implementation- and platform-aware models depend, respectively. This enables the investigation of the trade-offs discussed in Section~\ref{sec:intro} by working on models only, i.e., without requiring to compile and run QNNs on the target hardware.
\end{enumerate}



\section{Implementation-aware model}
This first phase takes in input a QONNX file representing a quantized neural network. Our framework expands the QONNX capabilities to allow a more detailed analysis of the memory and computation requirements of the QNN based on the implementation of a single operation. 
This step requires two inputs: \textbf{(1)} the QONNX representation of the QNN (it can be obtained from other model formats using state-of-the-art tools) and \textbf{(2)} an \textit{implementation configuration file}.
Listing~1 reports an example of the implementation configuration file.


The DAG representing the implementation-aware model can be formalized as $\mathcal{G} = (\mathcal{V}, \mathcal{E})$, where $\mathcal{V} = \{v_i \mid i = 1, ..., n\}$ denotes the set of $n$ nodes corresponding to operations, and $\mathcal{E} = \{e_{ij} \mid v_i, v_j \in \mathcal{V}\}$ represents the set of directed edges indicating data dependencies between operations. Each node $v_i \in \mathcal{V}$ is annotated with metadata such as the number of multiply-accumulate operations (MACs) and bit operations (BOPs) associated with the corresponding operation. Similarly, each edge $e_{ij} \in \mathcal{E}$ carries information about the amount of data, expressed in kilobytes (kB), produced by the source node $v_i$ and consumed by the destination node $v_j$, following the direction of the edge.

The implementation configuration file includes user-defined implementation details for each node (i.e., operation) in the QONNX model. 
Although in this work we focus on the main operations involved in Convolutional Neural Networks (CNN) inference, we adopt a modular approach that can be easily extended in future work to support other operations.
The considered operations are discussed next.

\begin{lstlisting}[style=pythonstyle, caption=Example of implementation file.]
    Quant_0:
        implementation: thresholds
        bit_width: 8
    MatMul_0:
        filter_wise: True
        implementation: LUT
        bit_width: 8
    Relu_0:
      implementation: comparator
    ...
\end{lstlisting}\label{li:impl_file}

\subsection{Convolutional layers}

Convolutional layers are the core components of CNN. Convolution applies a set of filters, also called kernels, which slide across the input feature maps to extract meaningful patterns such as edges or textures. Optimizing the execution of convolutional layers is hence crucial, particularly on resource-constrained platforms.

\begin{figure}[h!]
    \centering
    \includegraphics[width=1\linewidth]{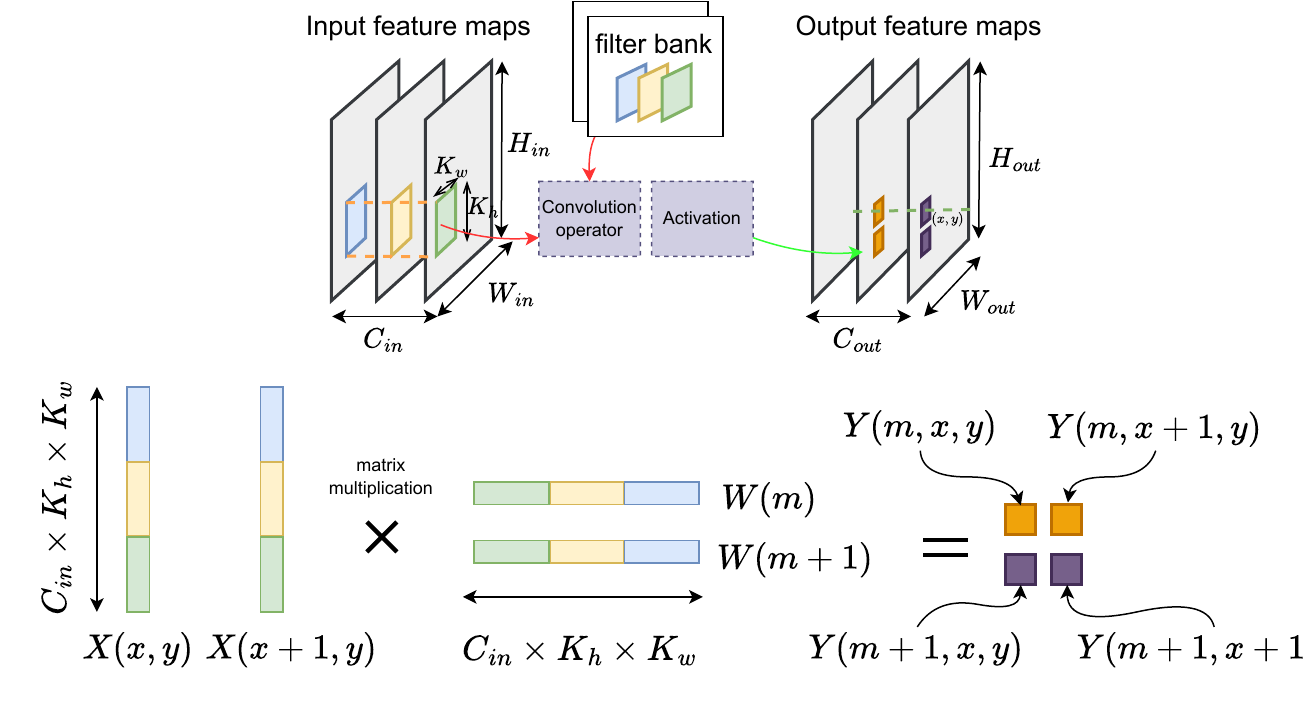}
    \caption{Comparison of the standard 2D convolution (top) and its equivalent implementation via \textit{im2col} transformation (bottom).}
    \label{fig:conv_vs_im2col}
\end{figure}

Consider an input tensor with shape $C_{in} \times H_{in} \times W_{in}$, where $C_{in}$ is the number of input channels, and $H_{in}$ and $W_{in}$ are the height and width of the input features, respectively. Each convolutional filter has size $C_{in} \times k_h \times k_w$, where $k_h$ and $k_w$ are the height and width of the kernel, respectively. A standard convolution computes each output value as a weighted sum between the filter and a local patch of the input, and repeats this across the entire spatial domain to generate an output tensor of size $C_{out} \times H_{out} \times W_{out}$, where $C_{out}$ is the number of filters, and $H_{out}$ and $W_{out}$ are respectively the height and width of the output tensor. An example convolution is illustrated in \cref{fig:conv_vs_im2col}.

This work considers a well-known optimization to reduce latency of convolutions, named \textit {im2col} transformation~\cite{im2col}, which converts a convolution into an efficient matrix multiplication. This is particularly effective on embedded devices that lack specialized convolution hardware but support optimized matrix operations~\cite{CMSIS-nn, pulp-im2col}. Instead of computing each convolution directly, the \textit{im2col} method rearranges the input such that each patch of the input that a kernel would process is unrolled into a column vector. This results in a buffer of size $(C_{in} \cdot k_h \cdot k_w) \times (H_{out} \cdot W_{out})$, where each column corresponds to one receptive field in the input. Note that, although the term unrolled is used, this does not mean flattening into a one-dimensional array, but rather organizing the data as a matrix suitable for multiplication.

Furthermore, the filters are reshaped into a matrix of shape $C_{out} \times (C_{in} \cdot k_h \cdot k_w)$, where each row corresponds to one filter. The convolution then becomes a matrix multiplication between the reshaped filters and the unrolled input buffer, producing a result of size $C_{out} \times (H_{out} \cdot W_{out})$, which can be reshaped back into the original output tensor format (see \ref{fig:conv_vs_im2col}). This reformulation allows the use of fast matrix multiplication routines, significantly accelerating execution. However, this comes at the cost of increased memory usage.

Fig.~\ref{fig:conv_vs_im2col} illustrates this transformation. The computation of a single output pixel in channel $m$ and position $(x, y)$ is expressed as $Y(m, x, y) = W(m) \cdot X(x, y)$,
where $W(m)$ is the $m$-th filter flattened into a row vector of length $C_{in} \cdot k_h \cdot k_w$, and $X(x, y)$ is the unrolled input patch centered at $(x, y)$, represented as a column vector of the same length. We assume the height-width-channel (HWC) data layout for tensors, since it showed~ \cite{CMSIS-nn} to provide better performance than the channel-height-width (CHW) format in most embedded use cases.

It is possible to introduce further data redundancy to boost the computation of the Convolution adopting a LUT, as described in Sec.~\ref{subsec:backgroundquantized}. In this way, the number of MAC operations needed is zero, while we increase the number of parameters needed.

\noindent
\textbf{Model decoration.} A \textit{Conv} node implemented using the im2col transformation lead to the following extension of the implementation DAG:
\begin{itemize}
    \item The memory required by the input tensor must account for the redundancy introduced by the im2col conversion:

    \begin{equation}\label{eq:input_mem_im2col}
        \text{Input memory} = (H_{\text{out}} \cdot W_{\text{out}})(C_{\text{in}} \cdot k_h \cdot k_w) \cdot L_x,
    \end{equation}

    where $L_x$ denotes the bit-width of the input features.
    
    \item Similarly, the memory requirements for the parameter and output edges are computed as:

    \begin{equation}\label{eq:param_mem_im2col}
        \text{Parameter memory} = (C_{\text{out}} \cdot C_{\text{in}} \cdot k_h \cdot k_w)
        \cdot 
        L_w + C_{\text{out}} \cdot L_{\text{acc}},
    \end{equation}
    \begin{equation}\label{eq:output_mem_im2col}
        \text{Output memory} = (C_{\text{out}} \cdot {H_{\text{out}} \cdot W_{\text{out}}}) 
        \cdot 
        L_{\text{acc}},
    \end{equation}
    
    where $L_{\text{acc}}$ is the bit-width of the accumulator (used for output precision), and $L_w$ is the bit-width of the weights.

    \item The operation node is renamed to \texttt{MatMul}, since the underlying operation is now a standard matrix multiplication. The node attributes are extended to include the number of MACs and BOPs, calculated as:

    \begin{equation}\label{eq:mac_im2col}
        \text{MACs} = C_{\text{out}} \cdot C_{\text{in}} \cdot k_h \cdot k_w,
    \end{equation}
    \begin{equation}\label{eq:bops_im2col}
        \text{BOPs} = \text{MACs} 
        \cdot 
        (1 + L_{\text{acc}} + L_w + L_x).
    \end{equation}
\end{itemize}
If, instead of using MAC operations, a look-up table (LUT) is adopted (i.e., $\text{MACs} = 0$), the number of parameters must be increased to include the size of the LUT, which is computed using Equation~(\ref{eq:lut-quant}), introduced later. The number of BOPs remains unchanged, since the MAC operation is replaced by a memory access operation indexed by the operands.

\subsection{Fully-connected layers}
Fully-connected layers consist of a matrix to vector multiplication, which can be seen as a special case of convolution where the input is treated as a single spatial location and the filter covers the entire input. Since there is no sliding of the filter and the input and weight dimensions already match, there is no need for \textit{im2col} transformations, since the dimension of the filters is the same as the input feature.

\noindent
\textbf{Model decoration.} It can be considered as a special case of the im2col implementation; hence, we use the decorations introduced for \textit{Conv}.

\subsection{Quantization}
The implementation of quantization operations strongly depends on the adopted quantization scheme. Indeed, for non-uniform quantization, except under special cases such as those in \cite{apot}, the re-quantization operation is accomplished by comparators arranged in a balanced tree so, where each threshold has the same precision of the input features, while the output is mapped in one of $2^b -1$ possible outputs, where $b$ is the expected precision of the output. Even if this approach is particularly fast, especially for low-bit quantization, it comes at the cost of a higher memory footprint. 

If the target platform does not present significant memory constraints, the quantization function can be implemented using a Look-Up Table (LUT), where each possible value of the input is directly mapped to its quantized counterpart. This can significantly accelerate computation, as the complexity of the operation is reduced to $\mathcal{O}(1)$ compared to the $\mathcal{O}(\log n)$ complexity of a balanced tree-based implementation. However, this improvement in speed comes at the cost of increased memory usage. The memory footprint of the LUT is:
\begin{equation}\label{eq:lut-quant}
    \text{Memory} = 2^{L_{\text{acc}}} \cdot L_{y},
\end{equation}

where $L_{\text{acc}}$ is the bit-width of the accumulator output from the previous layer, and $L_{y}$ is the target bit-width of the quantized output. It is important to note that this approach is not applicable when the input is in floating-point format, as the range of possible input values is not discrete and bounded in the same way as for integer accumulators.

Conversely, for uniform quantization we can adopt the \textit{dyadic scaling}. dyadic scaling is used to avoid the division operation, and it is common in practice~\cite{jacob2018quantization}, especially when we want to scale back the values of the accumulator to a lower precision integer. It approximates  the scale $S$ (recall \cref{subsec:backgroundquantized}) with two values $M$ and $n$ as 
    $S \approx m = \frac{M}{2^n},$
where $n$ is a positive integer lower than the highest precision on the considered platform, usually set to $30$ or $31$, while $M$ is a positive integer that can be computed offline in such a way $m$ closely approximates $S$. The computation of the scale can be implemented using multiplication and right bit-shifting, which are less expensive than division.

The adoption of dyadic scaling is suggested in the presence of constrained memory space, where the memory overhead due to the need to store thresholds is not acceptable. Indeed, it consists only of storing one parameter of the same precision as the accumulator. However, the savings in memory come at the cost of possible performance degradation, since we are approximating the scale $S$, which is usually represented as \texttt{FP32}, to the closest power-of-two integer. The difference between the original and the approximated is the error, which will propagate through the QNN during inference.

\noindent
\textbf{Model decoration.} A \textit{Quant} node implemented using the threshold-tree transformation leads to the following extension of the implementation DAG:
\begin{itemize}
    \item The memory requirements for the input and output edges can be computed using Eq.~(\ref{eq:input_mem_im2col}) and~(\ref{eq:output_mem_im2col}), adapted if the preceding node is a \texttt{Gemm} or \texttt{MatMul} operation.

    \item The memory requirements for the parameters depend on the chosen implementation:
    \begin{itemize}
        \item For \textbf{dyadic scaling}, it is limited to the 32-bits required for storing the scale parameter.

        \item For the \textbf{LUT-based implementation}, memory can be estimated using Equation~(\ref{eq:lut-quant}).

        \item For the \textbf{threshold-tree implementation}, the parameter memory is computed as:
        \begin{equation}\label{eq:th-tree}
            \text{Parameter memory} = (2^{L_y} - 1) \cdot L_{\text{acc}},
        \end{equation}
        where $L_{\text{acc}}$ is the bit-width of the accumulator (i.e., input precision) and $L_y$ is the bit-width of the output.
        
        \textbf{Note:} In the case of channel-wise quantization, the total parameter memory must be multiplied by the number of channels.
    \end{itemize}
    \item Node attributes are extended to include the number of BOPs, calculated based on the implementation method:

    \begin{itemize}
        \item For the \textbf{threshold-tree} implementation:
        \begin{equation}\label{eq:bops_threshold}
            \text{BOPs}_{\text{threshold}} = {I} 
            \cdot 
            \log_2(T) \cdot L_{\text{acc}},
        \end{equation}
        where $I$ is the number of input features, $T$ is the number of thresholds computed as $T = 2^{L_y} - 1$, and $L_{\text{acc}}$ is the bit-width of the input (accumulator).
    
        \item For the \textbf{dyadic scaling} implementation:
        \begin{equation}\label{eq:bops_scaling}
            \text{BOPs}_{\text{scaling}} = I
            \cdot 
            \texttt{\#bit-shifts},
        \end{equation}
        where $\texttt{\#bit-shifts}$ indicates the number of shift operations applied during the quantization.
    \end{itemize}
\end{itemize}

\subsection{Activation functions}
Usually, the non-linear activation function adopted by QNN deployed on constrained devices is the ReLU function, due to its easy implementation via comparators between the input features and zero. However, more complex activation functions can also be adopted, leveraging the same threshold-tree previously introduced for quantization operations. Formally, a set of $T$ thresholds $\{\theta_1, \theta_2, \ldots, \theta_T\} \in \mathbb{Z}$, each represented with the same bit-width as the input features, is used to discretize the activation function into \( T+1 \) output levels. Also in this case, each threshold has the same bit-width as the input features, but in this case, the number of thresholds is a user-defined parameter. Indeed, the adoption of thresholds corresponds to the conversion of the activation function to a step function, therefore, the higher  is the number of thresholds $T$ the closer is the activation function is with respect to the original one.
However, more thresholds also correspond to an higher number of parameters to store in memory for the computation.
Ancillary operations of this kind may also take advantage specialized hardware extensions, such as the one for DSP of RISC-V cores \cite{dory}.

\noindent
\textbf{Model decoration.} We considered the ReLU function as activation function due to its effectiveness and hardware-friendly implementation via comparators. An \textit{Act} node implemented using the ReLU function via a comparator leads to the following extension of the implementation DAG:

\begin{itemize}
    \item The memory requirements for the input and output edges can be computed using Eqs.~(\ref{eq:input_mem_im2col}) and~(\ref{eq:output_mem_im2col}), adapted if the preceding node is a \texttt{Gemm} or \texttt{MatMul} operation.
    
    \item Node attributes are extended to include the number of BOPs, computed as:
    \begin{equation}\label{eq:bops_relu}
        \text{BOPs} = I 
        \cdot 
        (L_x + 1),
    \end{equation}
    where $I$ is the number of input features and $L_x$ is the bit-width of the input features.
\end{itemize}

\subsection{Pooling operations}
The implementation of pooling operations, such as \textit{max pooling} and \textit{average pooling}, are relatively easier.
For the max pooling, for each patch of the input feature,  we compute the maximum leveraging comparators between the patch values. The average pooling is, in principle, more difficult, since it should require a division by the number of elements in the patch. Since the division is an expensive operation in hardware, we approximate it as we explained for the dyadic scaling in the closest division to a power of two, i.e., leveraging the much less expensive bit-shifting operation.

\noindent
\textbf{Model decoration.} We currently evaluated only the Max Pool function as the pooling operation; however, the extension to other pooling functions is straightforward. A \textit{Max Pool} node implemented using comparators leads to the following extension of the implementation DAG:

\begin{itemize}
    \item The memory requirements for the input and output edges are computed using Eq.~(\ref{eq:input_mem_im2col}) and~(\ref{eq:output_mem_im2col}), adapted appropriately if the preceding node is a \texttt{Gemm} or \texttt{MatMul} operation.
    
    \item Node attributes are extended to include the number of BOPs, computed as:
    \begin{equation}\label{eq:bops_maxpool}
        \text{BOPs} = I 
        \cdot 
        (L_x \cdot K_w \cdot K_h),
    \end{equation}
    where $I$ is the number of inputs, $L_x$ is the bit-width of the input features, and $K_w$ and $K_h$ are the horizontal and vertical dimensions of the pooling kernel, respectively.
\end{itemize}


\section{Platform-aware model and scheduling}

In the final phase, we include information about the target deployment platform and the execution time bound of operations. This allows bounding the overall latency required to complete one inference pass. In this way, ALADIN can connect interference time with implementation choices and quantization precision, providing valuable insights and eliminating the need to deploy the QNN on the physical device during early development stages, thus significantly reducing design and development times.

Furthermore, if the target platform is reconfigurable—as is the case with many modern AI accelerators~\cite{operand}—this approach enables hardware-software co-design, i.e., it allows also exploring how the amount of hardware resources affect latency and accuracy of feasible deployments. ALADIN can indeed identify potential bottlenecks in the architecture, which can then be addressed by reallocating hardware resources more effectively.

\noindent
\textbf{Producing the platform-aware model.}
The implementation aware model is refined by splitting operations into sub-operations that have sufficient granularity to be individually scheduled on the target hardware. The data required to accomplish operations are split as well into chunks based on either the output features or channels, e.g., the output features of a matrix multiplication can be computed by fixing the input feature and iterate over the relative parameters. All such model transformations depend on the hardware configuration, namely the number of parallel cores, memories, and memory sizes.

\noindent
\textbf{Scheduling.}
Determining the optimal schedule for a QNN is beyond the scope of this paper and left as future work. As such, we opted to rely on Dory~\cite{dory}, a state-of-the-art automatic tool to deploy QNNs on embedded platforms. In Dory, data are classified into four main categories: input data, output data, parameters, and temporary buffers. While the first three ones are self-explanatory, temporary buffers are used to store auxiliary structures required for layer execution, such as the LUT tables or threshold-trees described in the previous section. Because data movement between memories is costly, Dory directly allocates these auxiliary structures in the L1 buffer, similarly to output data, storing them on-chip to minimize transfers.

In terms of scheduling, when all the required data for a given layer fit entirely within the L1 memory, no data tiling is needed, and the layer can be executed in a single pass. Otherwise, Dory partitions the data based on the output channels or feature maps to ensure that each tile fits within the available L1 space. If memory utilization allows, Dory can also employ a \textit{double-buffering} strategy, which reserves twice the space of a single buffer but enables overlapping of data transfer and computation. This prefetching mechanism effectively hides the latency of DMA transfers, both between L3 and L2 and between L2 and L1 memories, thereby reducing the overall execution time.

\section{Evaluation}
\label{sec:exp}
We now evaluate the capabilities of the proposed design flow by analyzing a well-known compact CNN, MobileNet V1~\cite{mv1}, under different mixed-precision and implementation configurations. The CNN is trained on the CIFAR-10 dataset~\cite{cifar}, and it is composed by several 2D convolutional layers and on fully connected layers, as classification head. After a full-precision training step the model was quantized via quantization-aware training (QAT) using Brevitas~\cite{brevitas}, where batch normalization layers are handled by folding them in the previous convolutional layer \cite{bn_folding}. The evaluation focused on the GAP8~\cite{GAP8} platform.

We compare three different mixed-quantization and implementation configurations (\textbf{Case 1-3}), reported in Table~\ref{tab:merged_config_mnv1}, with the corresponding accuracy. 
  We focus on byte and sub-byte precisions (2, 4, and 8 bits). The accumulators used for  intermediate results are 32-bits, except in sub-byte quantization configurations, where 16-bit ones are used.
The code is available at the following link: \url{https://github.com/balditommaso/ALADIN}




\begin{table}[t!]
\caption{Quantization precision (bitwidth) and implementation configuration of three different MobileNetV1 instances. Each block includes a depthwise and a standard convolution, both followed by ReLU and quantization layers. Average pooling is applied before the classifier.}
\label{tab:merged_config_mnv1}
\centering
\small
\setlength{\tabcolsep}{3pt}        
\renewcommand{\arraystretch}{1.05}
\begin{tabular}{|c|cc|cc|cc|}
\hline
\rowcolor{lightgray}
\textbf{Block} 
& \multicolumn{2}{c|}{\textbf{Case 1}} 
& \multicolumn{2}{c|}{\textbf{Case 2}} 
& \multicolumn{2}{c|}{\textbf{Case 3}} \\
\cline{2-7}
\rowcolor{lightgray}
& \textbf{Prec.} & \textbf{Impl.} 
& \textbf{Prec.} & \textbf{Impl.} 
& \textbf{Prec.} & \textbf{Impl.} \\
\hline
Pilot      & int8 & im2col & int8 & im2col & int8 & im2col \\
\hline
Block$_1$    & int8 & im2col & int4 & im2col & int8 & im2col \\
\hline
Block$_2$    & int8 & im2col & int4 & im2col & int4 & im2col \\
\hline
Block$_3$    & int8 & im2col & int4 & im2col & int4 & im2col \\
\hline
Block$_4$    & int8 & im2col & int4 & im2col & int4 & im2col \\
\hline
Block$_5$    & int8 & im2col & int4 & im2col & int4 & im2col \\
\hline
Block$_6$    & int8 & im2col & int4 & im2col & int4 & LUT \\
\hline
Block$_7$    & int8 & im2col & int4 & im2col & int4 & LUT \\
\hline
Block$_8$    & int8 & im2col & int4 & LUT    & int4 & LUT \\
\hline
Block$_9$    & int8 & im2col & int4 & LUT    & int4 & LUT \\
\hline
Block$_{10}$ & int8 & im2col & int4 & LUT    & int2 & LUT \\
\hline
Classifier & int8 & Gemm   & int8 & Gemm   & int4 & LUT \\
\hline
\rowcolor{lightgray}
\textbf{\textit{Accuracy}} 
& \multicolumn{2}{c|}{0.83} 
& \multicolumn{2}{c|}{0.77} 
& \multicolumn{2}{c|}{0.78} \\
\hline
\end{tabular}
\end{table}


\subsection{QNN implementation analysis}
Our approach is capable of providing a first set of quantitative data on the computational and memory costs of each layer by leveraging the implementation DAG in a platform-independent manner (see Section~\ref{sec:method}).

\begin{figure}[]
    \centering
    \subfloat[]{\includegraphics[width=1\linewidth]{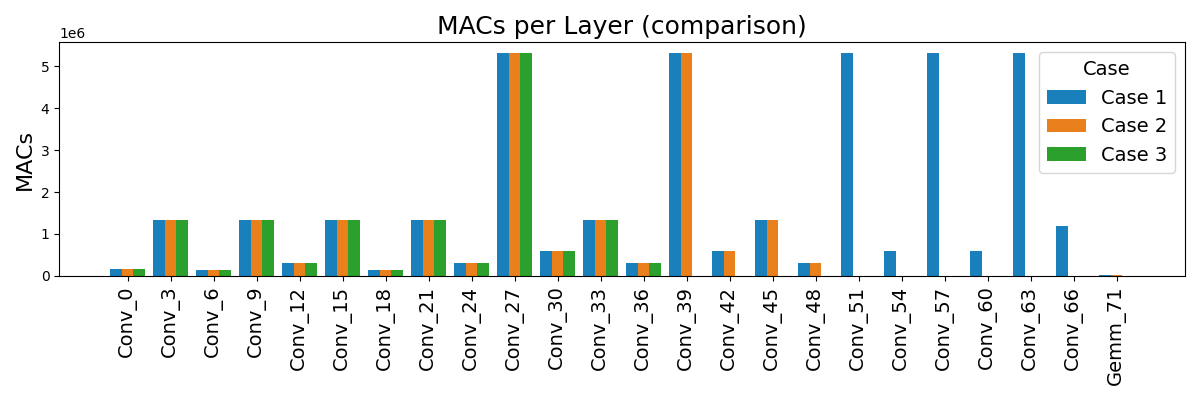}\label{subfig:mac}}\\
    \subfloat[]{\includegraphics[width=1\linewidth]{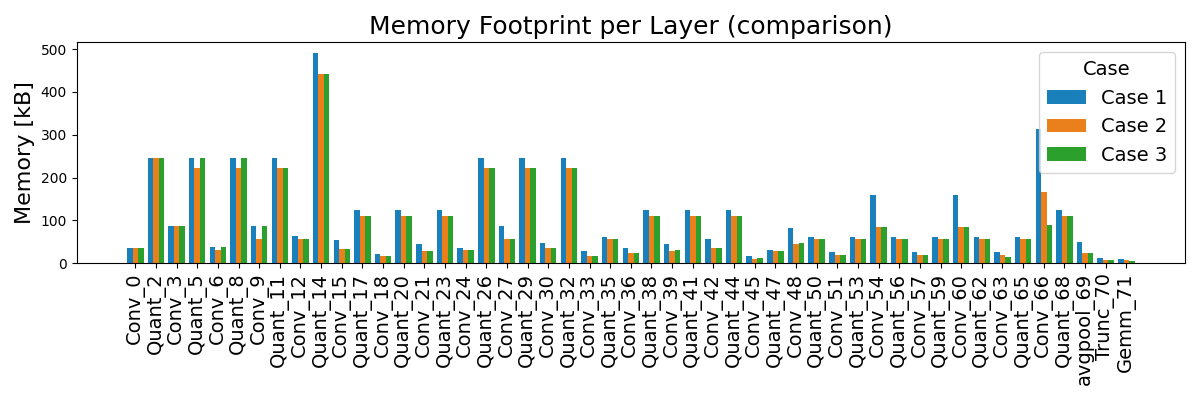}\label{subfig:memory}}\\
    \subfloat[]{\includegraphics[width=1\linewidth]{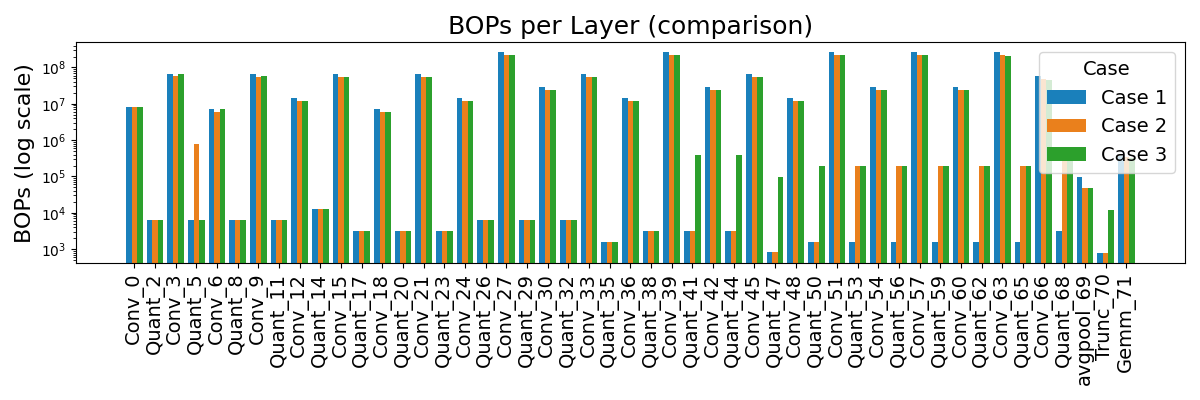}\label{subfig:bops}}
    \caption{Each plot represents a different metric under analysis: (a) compares the network layer-wise in terms of MACs (irrelevant nodes are excluded); (b) shows the memory footprint of each layer; and (c) reports the complexity of each layer in terms of BOPs. ReLU layers are omitted, as their implementation does not vary among the different configurations.}
    \label{fig:model_load}
\end{figure}
We compared the metrics extracted from the analyzed model, shown in Figure~\ref{fig:model_load}, and we observed the following.

\noindent    \textbf{Depthwise vs. standard convolutions}\footnote{A standard convolution applies a unique 3D kernel to all input channels simultaneously and sums the results to produce each output channel, mixing both spatial and cross-channel information.
    In contrast, a depthwise~\cite{depthwise} convolution applies a separate 2D filter to each input channel independently, capturing only spatial features without combining information across channels.}: By looking at the results for $\text{Block}_{10}$, which includes the depthwise and standard convolution layers of the block and the associated quantization layers \texttt{Quant\_65} and \texttt{Quant\_68}, it emerges that depthwise convolutions are more computation-intensive in terms of MAC operations (Figure~\ref{subfig:mac}) compared to standard convolutions. However, they exhibit a substantially lower memory footprint (Figure~\ref{subfig:memory}). This makes depthwise convolutions particularly suitable for MAC reduction via LUT-based multiplications. Even for higher bit-widths, the memory footprint remains contained, whereas in standard convolutions, such as \texttt{Conv\_66} and \texttt{Conv\_42}, variations in bit-width and implementation significantly affect the overall memory overhead.

 \noindent    \textbf{Low-bit quantization and threshold-tree}: When analyzing quantization layers in terms of memory footprint and computational overhead (expressed in BOPs; see Figures~\ref{subfig:memory} and~\ref{subfig:bops}), we observe that threshold-tree implementations, even under low-bit quantization, introduce a memory overhead comparable to 8-bit quantization based on dyadic scaling. Therefore, this approach should be adopted judiciously, only when the target platform can provide a significant speed-up compensating for this cost.

These analyses, performed in a platform-independent manner,  provide valuable quantitative insights into computational and memory bottlenecks.

\begin{figure}[]
    \centering
    \subfloat[]{\includegraphics[width=1\linewidth]{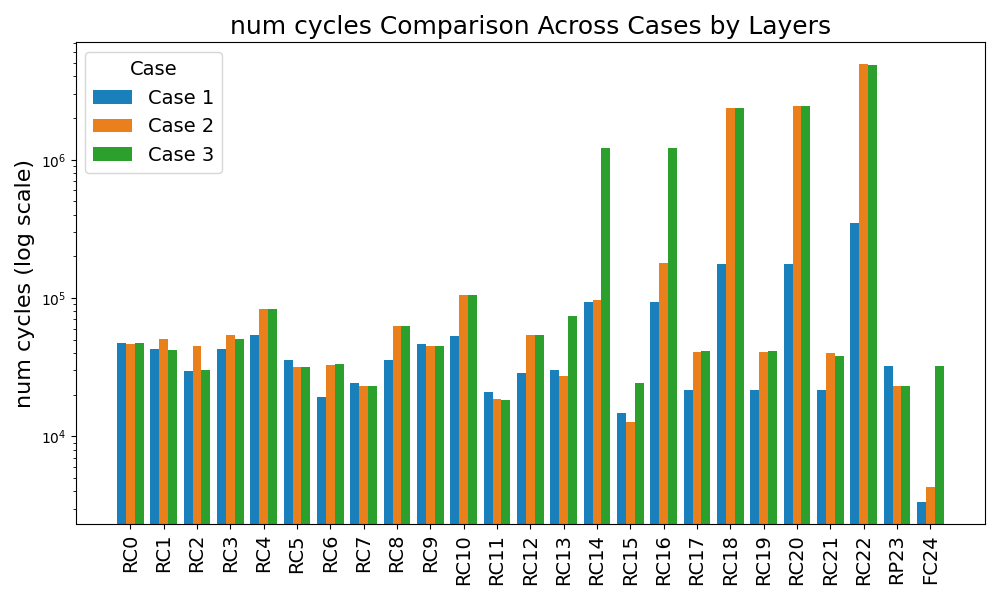}\label{subfig:cycles}}\\
    \subfloat[]{\includegraphics[width=1\linewidth]{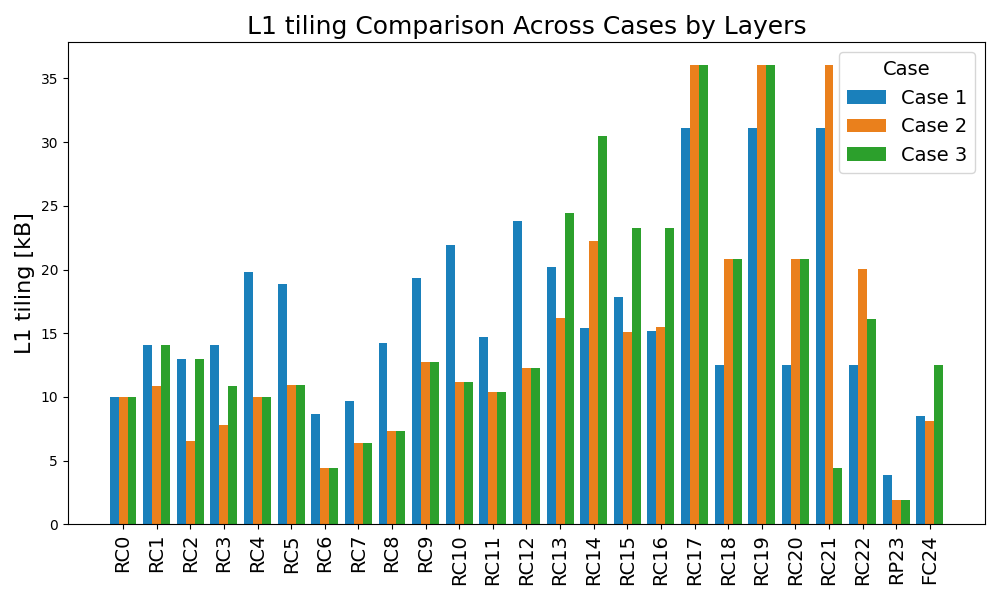}\label{subfig:l1-tiling}}\\
    \subfloat[]{\includegraphics[width=1\linewidth]{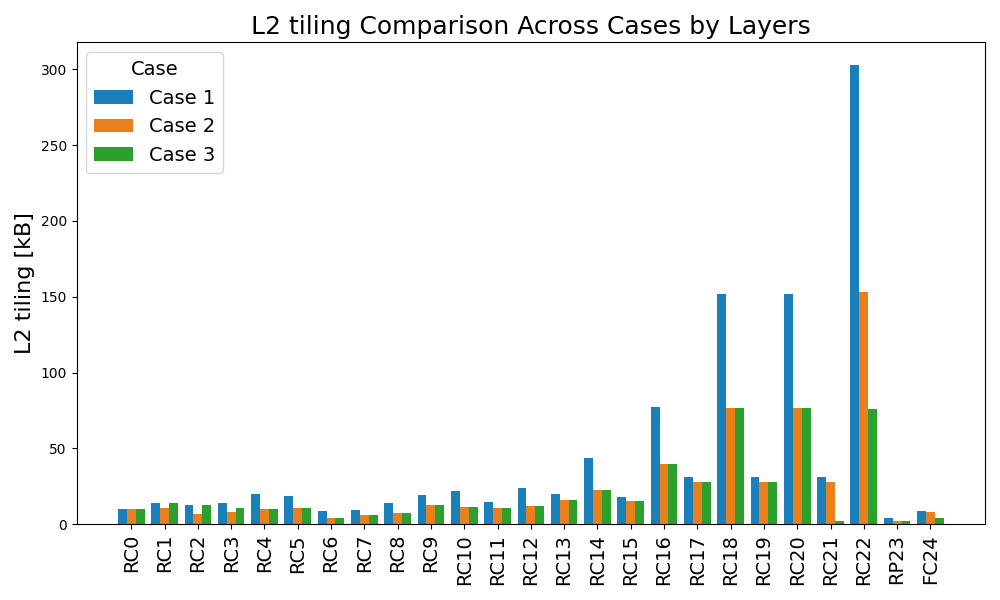}\label{subfig:l2-tiling}}
    \caption{Each plot represents a different performance metric under analysis: 
    (a) compares the networks layer-wise in terms of execution cycles, excluding non-relevant nodes; 
    (b) reports the L1 memory footprint derived from data produced and consumed by each layer; 
    and (c) shows the corresponding L2 memory utilization. ReLU layers are omitted since their implementation remains unchanged across configurations.}
    \label{fig:model_perf}
\end{figure}

\begin{figure*}[]
    \centering
    \subfloat[]{\includegraphics[width=0.48\linewidth]{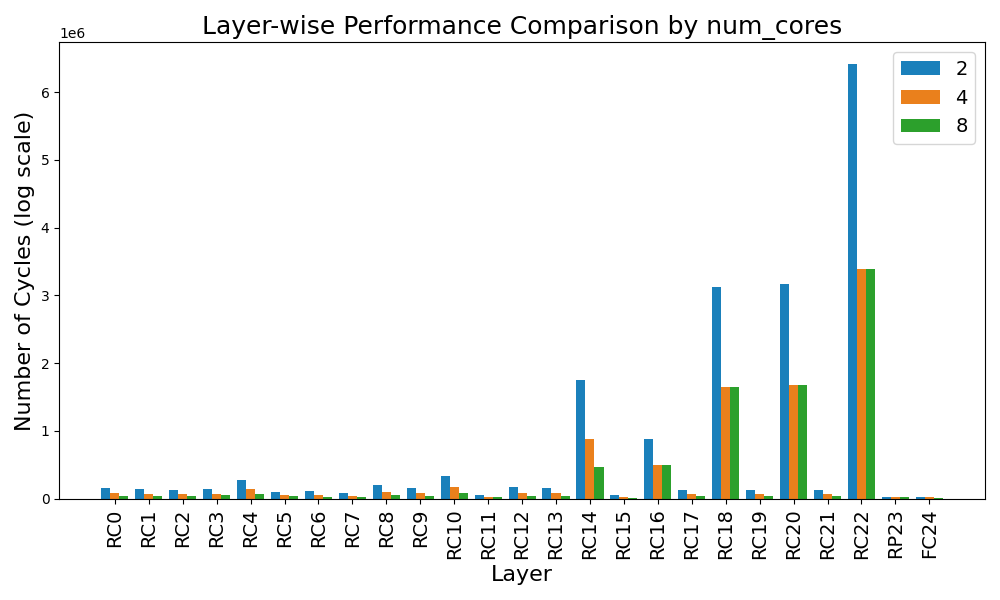}\label{subfig:cycle_by_core}}
    \hfill
    \subfloat[]{\includegraphics[width=0.48\linewidth]{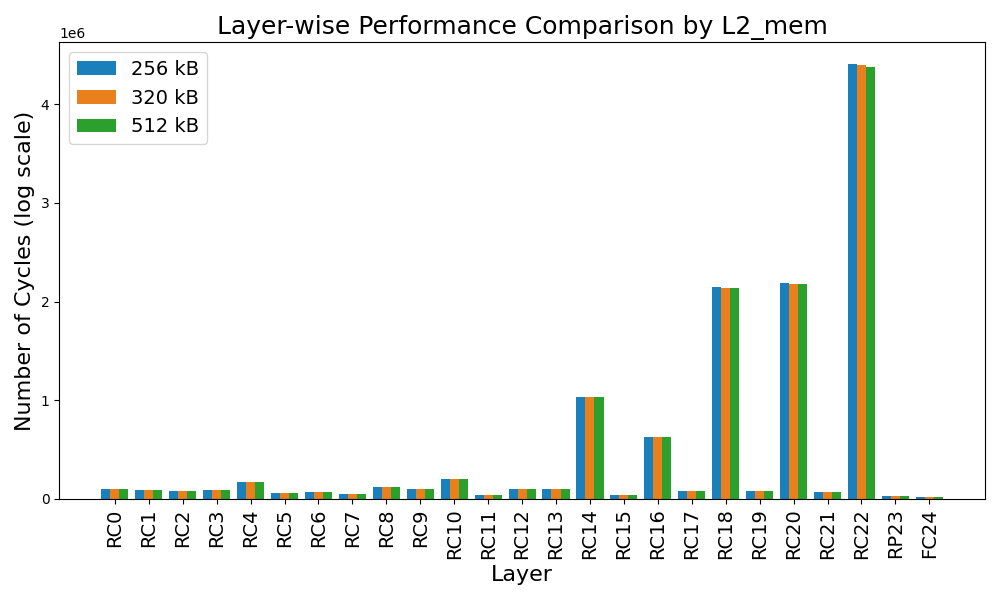}\label{subfig:cycle_by_l2}}\\[2mm]
    \subfloat[]{\includegraphics[width=0.48\linewidth]{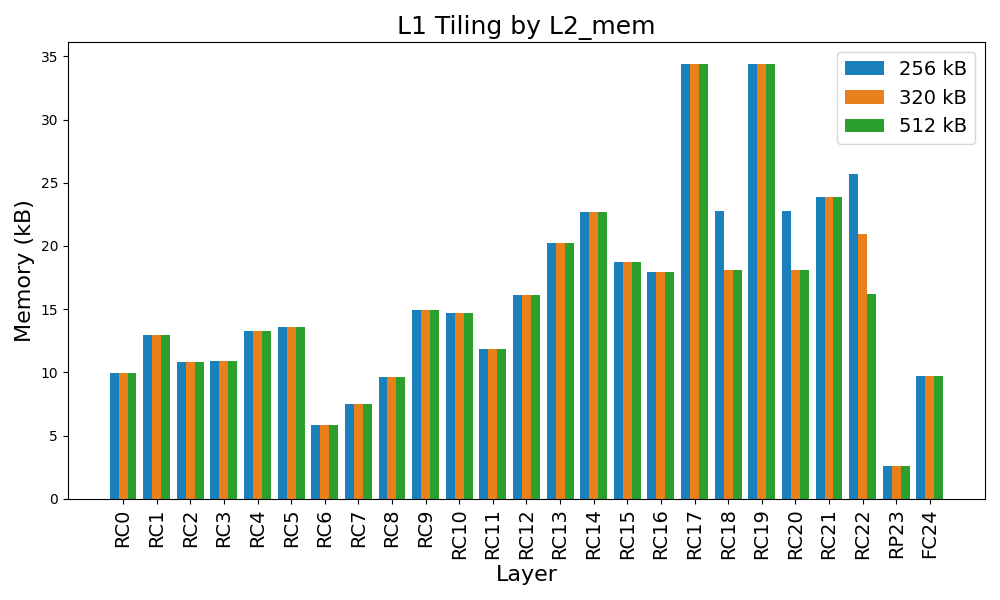}\label{subfig:l1_tiling_by_l2}}
    \hfill
    \subfloat[]{\includegraphics[width=0.48\linewidth]{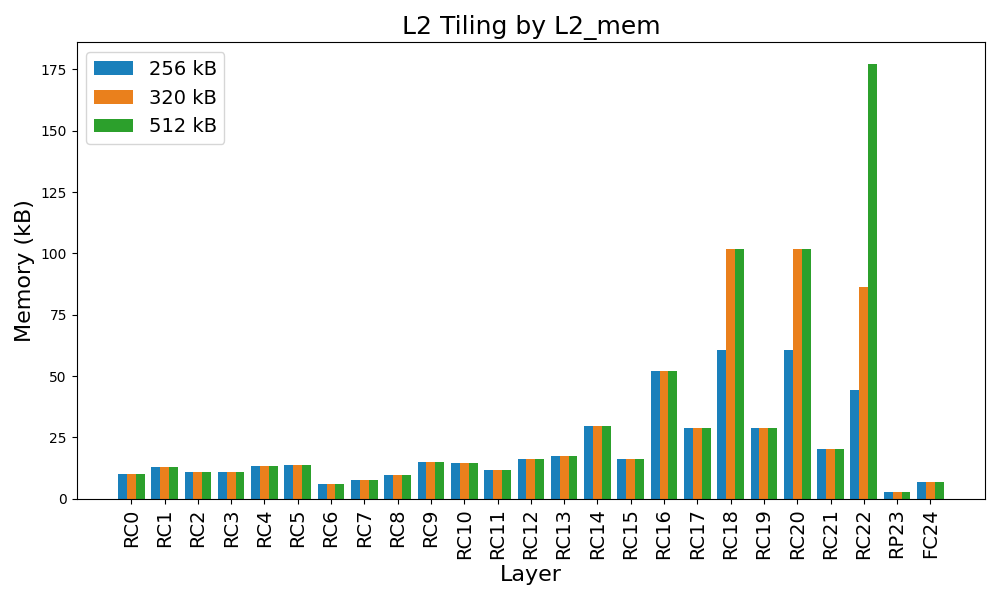}\label{subfig:l2_tiling_by_l2}}
    \caption{Model performance in terms of clock cycles (top row) as a function of the number of cores and L2 memory capacity. The bottom row reports the corresponding L1 and L2 tiling configurations under different L2 memory settings. RC = ReLU-Convolution, RP = ReLU-Pooling, FC = Fully Connected.}
    \label{fig:co-design}
\end{figure*}

\subsection{Performance analysis}

By incorporating platform-specific information, such as system capabilities and the execution cost of each basic operation, we can bound the inference time on a per-layer basis. This enables the identification of potential bottlenecks in the execution pipeline, which can be addressed either from an algorithmic perspective, by modifying the QNN quantization configuration or adjusting the scheduling strategy, or from a hardware perspective,  as discussed in a later Section~\ref{sec:hwexploration}.
The target platform for this experiment is the one described in Section~\ref{sec:platformmodel}, with a cluster of 8 cores, an L2 memory of 512~\textit{kB}, and an L1 memory organized into 16 banks of 64~\textit{kB} each.
The inference time bound obtained from the model is compared against the inference performance provided by GVSoC~\cite{gvsoc}, a cycle-accurate simulator of RISC-V-based platforms, such as GAP8. On top of that, we extended Dory's frontend, capable of generating C code that runs on the platform, from our platform-aware DAG, and we also contributed to the RISC-V DSP-optimized kernel library for GAP8~\cite{XpulpNN} by adding the possibility to perform multiplications using pre-computed LUTs. These extensions were required to demonstrate the capabilities of the proposed work,
even if the RISC-V-based cores of the 
GAP8's cluster are optimized to efficiently perform MAC-intensive operations, thus leading to a significant reduction in terms of clock cycles with respect to LUT-based implementations. Different target platforms, equipped with specific accelerators, such as \cite{lut2}, may instead highlight the speed-up coming from the adoption of LUT-based multiplications.

Figure~\ref{fig:model_perf} shows the performance of the model in terms of clock cycles and the memory utilization of both L1 and L2 SRAM. Dory applies operator fusion as an optimization technique to improve execution of the C code generated for the target platform; therefore, the layer shown in the following plots represents the operators resulting from fusing a convolution or a fully connected layer with ReLU and quantization. From the plots, we can observe the following.

\noindent \textbf{Im2col-based implementations}: A comparison between Figures~\ref{subfig:mac} and~\ref{subfig:cycles} highlights the coherence between the implementation-level and platform-level analyses. In the first blocks, where convolutions are implemented using the \textit{im2col} strategy, the number of cycles required for 4-bit convolutions is comparable to that of 8-bit ones. This behavior is primarily due to the bit-unpacking mechanism of the target platform, which introduces additional overhead. However, as shown in Figures~\ref{subfig:l1-tiling} and~\ref{subfig:l2-tiling}, the low-bit \textit{im2col} implementation exhibits a reduced memory footprint, enabling more efficient prefetching strategies that can partially hide the latency associated with DMA data transfers. Despite this advantage, the expected performance gain is limited in the initial layers of the network, which contain relatively few output channels,  thereby restricting parallelization opportunities. This observation is confirmed by analyzing \texttt{RC\_13} and \texttt{RC\_15} in Cases~1 and~2. These layers, implementing standard convolutions with a higher number of output channels, show a significant reduction in execution cycles (note that the scale is logarithmic), showing how increased parallelization efficiency, enabled by lower memory utilization, translates into improved performance.

    \noindent \textbf{
    LUT-based implementations}: Interesting observations can be made regarding the LUT-based implementations, particularly when comparing \texttt{RC\_21} and \texttt{RC\_22} across Cases~2 and~3. These two layers adopt LUTs parameterized with 4-bit and 2-bit weights, respectively. Recalling the analysis of LUT dimensionality discussed in Section~\ref{sec:lut}, and by comparing the memory footprint estimated in Figure~\ref{subfig:memory} with the L1 and L2 memory utilization shown in Figures~\ref{subfig:l1-tiling} and~\ref{subfig:l2-tiling}, we can observe a clear memory overhead for higher bit-widths, which can harm timing predictability. This is due to the exponential growth of LUT size with respect to weight precision. However, the expected speed-up from the 2-bit LUT implementation in Case~3 is not evident in the cycle count shown in Figure~\ref{subfig:cycles}. This unexpected behavior arises from the specific LUT management strategy adopted by the target platform, where the LUT is stored contiguously in L1 memory and shared among all cores in the cluster. The smaller LUT thus exhibits a higher level of concurrent access compared to the larger one, since multiple cores attempt to read from the same smaller address space, creating a bottleneck that limits the anticipated performance gain. Alternative architectures, such as those presented in~\cite{lut2}, mitigate this issue by replicating LUT instances to increase parallelism. However, this approach is not always feasible on memory-constrained platforms.


These complex trade-offs would have been difficult to detect and analyze without ALADIN. Furthermore, in real-time systems, our framework provides valuable support to system architects who are called to co-design complex AI workloads on constrained hardware while simultaneously satisfying deadline constraints, HW resource constraints, and maximizing accuracy.

\subsection{HW Design Evaluation}
\label{sec:hwexploration}

We now demonstrate the capability of the proposed tool to support hardware design evaluation for platforms with reconfigurable features. Specifically, GVSoC allows reconfiguration of the target platform by varying both the SRAM capacity and the number of cores within the cluster. 
For this evaluation, we use a fixed model configuration (\textit{Case~2}) while evaluating different hardware settings by modifying the number of cores and the L2 memory capacity. The L1 memory, on the other hand, cannot be increased beyond 64~$kB$, and significantly reducing its capacity results in schedulability failures, as expected by the implementation-aware model.
To validate this capability, we conducted a grid search, which serves as a proof of concept for the proposed framework. The explored design space includes the number of cores (\{2, 4, 8\}) and the L2 SRAM capacity (\{256, 320, 512\}~$kB$).
From Figure~\ref{fig:co-design}, we observe that for operations with low memory intensity, increasing the number of cores yields a clear performance boost, as seen in the initial convolutional blocks. More interestingly, for deeper standard convolution layers such as \texttt{RC\_18}, \texttt{RC\_20}, and \texttt{RC\_22}, which are both more memory-intensive and highly parallelizable, the performance gain in terms of clock cycles saturates beyond four cores. In these cases, further improvements can only be achieved by increasing the L2 memory capacity.
This behavior aligns with the findings discussed in the previous section: for memory-intensive layers, a larger L2 SRAM enables greater data reuse, reducing the need for costly DMA transfers between L3 and L2. At the same time, it decreases L1 SRAM utilization, allowing more effective prefetching strategies, mitigating the computational overhead of DMA transfers between L1 and L2. These effects collectively translate into lower execution times, as reflected in the reduction of total clock cycles.


\section{Conclusion}

This paper presents ALADIN, an accuracy-latency-aware design-space inference analysis framework for mixed-precision QNNs on resource-constrained platforms. Our approach allows for significantly reducing development time and iteration cost, uncovering computational and memory bottlenecks, and providing system designers of real-time systems with a practical tool to evaluate these trade-offs under tight real-time and resource constraints. Through  refining models from platform-independent to hardware-aware representations and validating them on a cycle-accurate simulator of a RISC-V-based AI platform, the proposed methodology supports effective hardware–software co-design and the development of real-time AI systems with tight memory and latency constraints. Future work will focus on deriving optimized scheduling strategies.

\bibliographystyle{IEEEtran}
\bibliography{bibliography}
\end{document}